\documentclass[letterpaper,prl,twocolumn,groupedaddress]{revtex4}[10pt] 
\usepackage{graphicx}	
\usepackage{dcolumn}
\usepackage{amssymb}	
\usepackage{amsmath}	
\usepackage{hyperref}
\usepackage{multirow}
\usepackage{array}
\usepackage{bm}
\usepackage{latexsym}
\usepackage{geometry}
\usepackage{color}	
\newcommand{\be}{\begin{equation}}
\newcommand{\ee}{\end{equation}}
\newcommand{\bea}{\begin{eqnarray}}
\newcommand{\eea}{\end{eqnarray}}

\def\corr#1{\textcolor{red}{#1}}	

\begin{document} 


\title{Spectrum and Wave Functions of Excited States in Lattice Gauge Theory}


 \author{
H.~Kr\"{o}ger$^{a,b}$\footnote{Corresponding author, Email:
hkroger@phy.ulaval.ca},
A.~Hosseinizadeh$^{a}$, 
J.F.~Laprise$^{a}$, 
J.~Kr\"{o}ger$^{c}$
} 

\affiliation{
$^{a}$ {\small\sl D\'{e}partement de Physique, Universit\'{e} Laval, Qu\'{e}bec, Qu\'{e}bec G1V 0A6, Canada} \\
$^{b}$ {\small\sl Frankfurt Institute for Advanced Studies, Goethe Universit\"at Frankfurt, 60438 Frankfurt am Main, Germany\footnote{present address}} 
\\
$^{c}$ {\small\sl Physics Departement and Center for the Physics of Materials, McGill University, 
Montr\'{e}al,  Qu\'{e}bec H3A 2T8, Canada} 
}

\date{\today, \corr{(Corrections in red)}}


\begin{abstract}
We suggest a new method to compute the spectrum
and wave functions of excited states. We construct a stochastic basis of Bargmann link states, drawn from a physical probability density distribution and compute transition amplitudes between stochastic basis states. From such transition matrix we extract wave functions and the energy spectrum. We apply this method to $U(1)_{2+1}$ lattice gauge theory. As a test we compute the energy spectrum, wave functions and thermodynamical functions of the electric Hamiltonian and compare it with analytical results. We find excellent agreement. We observe scaling of energies and wave functions in the variable of time. We also present first results on a small lattice for the full Hamiltonian including the magnetic term.
\end{abstract}


\maketitle


\section{Introduction}
\label{sec:Intro} 
Much progress has been made in lattice gauge theory on the computation of hadron spectra \cite{Gattringer07,Frigori07,Lang07,Gattringer08,Cais08,Bulava07,Walker08}.
Such spectra can be extracted from a matrix of 2-point correlation functions
built from extended operators with suitable quantum numbers applied to the ground state \cite{Basak05}. 
Here we present a new approach for the computation of spectra and wave functions suggesting
to compute transition matrix elements between states chosen from a stochastic basis with respect to gauge degrees of freedom.
We consider transition amplitudes
\begin{equation}
\label{eq:TransAmplUpsilon}
M_{\mu,\nu}(T) = 
\langle \Upsilon_{\nu} | \exp[-H T/\hbar ] | \Upsilon_{\mu} \rangle ~ , ~ \mu, \nu=1,\dots,N ~ ,
\end{equation}
where $|\Upsilon_{\nu} \rangle$ denotes a time-independent Bargmann-link state, i.e., a configuration of link variables $U_{ij}$ assigned to all of the links $ij$ on the {\it spatial} lattice. 
It is crucial to choose states $|\Upsilon_{\nu} \rangle$ which are physically relevant and important. 
We use as stochastic technique Monte Carlo with importance sampling to sample states from a large variety of possibilities. The stochastic basis states are closely related to equilibrium path configurations in the Euclidean path integral. 
From such matrix $M_{\mu,\nu}(T)$ 
we extract a spectrum and wave functions of an effective Hamiltonian - the 
so called Monte Carlo Hamiltonian - being valid in a low energy, respectively 
low-temperature window. The Monte Carlo Hamiltonian has been 
suggested 1999~\cite{Jirari99}. In field theory, the Monte Carlo Hamiltonian 
has been applied to the $1+1$ Klein-Gordon model for the computation of the spectrum and 
thermodynamical functions~\cite{Caron01,Luo01b,Luo01c,Kroger03b}, and 
likewise to the $1+1$ scalar model for the computation of the spectrum and 
thermodynamical functions~\cite{Huang02,Kroger03a,Kroger07}. A first step towards 
the Monte Carlo Hamiltonian in lattice gauge theory has been made in~\cite{Paradis07} 
by computing transition amplitudes of $U(1)$ gauge theory.
Here we construct the Monte Carlo Hamiltonian for $U(1)_{2+1}$ lattice gauge 
theory and apply it to compute the energy spectrum of excited states, the corresponding 
wave functions and thermodynamical functions. Here we want to show the 
working of the method applied to an Abelian, but non-trivial model in lattice gauge 
theory. In the case of the electric Hamiltonian we consider spatial lattice volumes 
up to $10^{2}$, while for the full Hamiltonan we present only preliminary results on a $2^{2}$ lattice.
Because of the smallness of the lattices we do not consider here the quantum continuum 
limit ($a \to 0$). Rather, we try to give a careful analysis of the origin and size 
of errors, which determine the limitations of the method. 
\begin{figure}[h,t]
\centering
\includegraphics[width=0.49\textwidth]{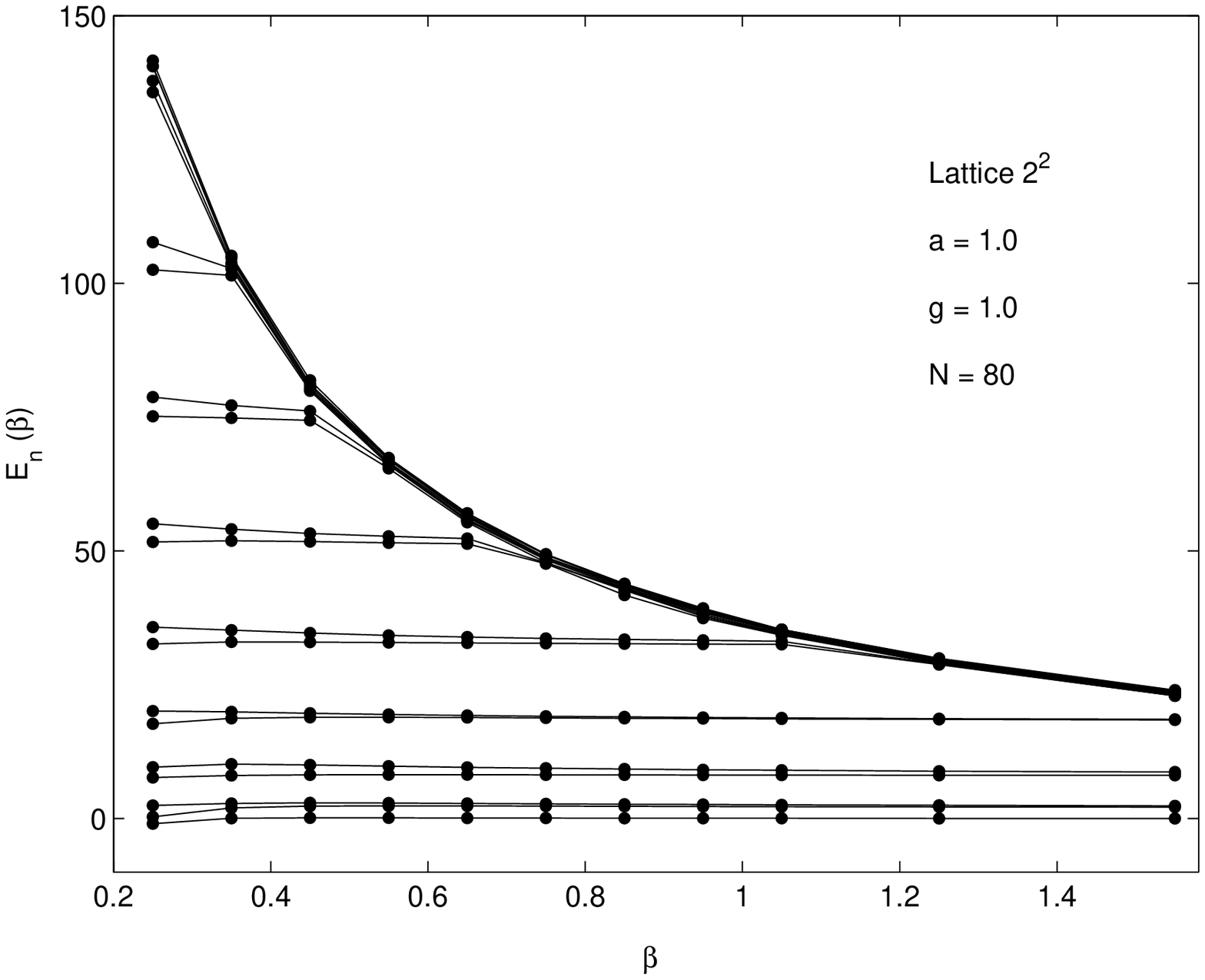} 
\includegraphics[width=0.49\textwidth]{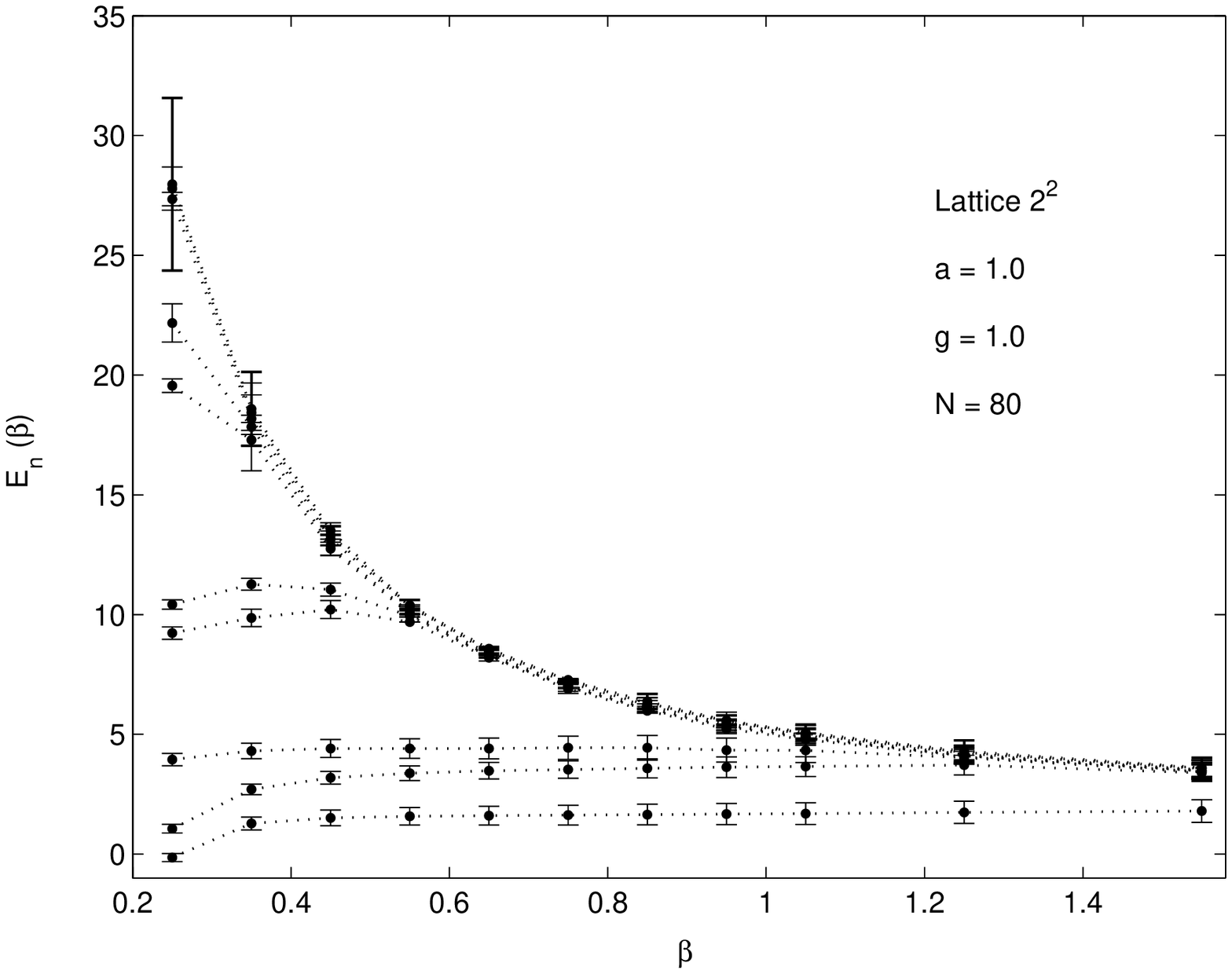}
\caption{Scaling of energy levels from the Monte Carlo Hamiltonian. Comparison of electric Hamiltonian (left) with full Hamiltonian (right). The bars represent statistical errors. 
$2^{2}$ spatial lattice, $a_{s}=a_{t}=1$, $g=1$, $N=80$.}
\label{fig:ScalingCompareElec+FullHam}
\end{figure}

\section{$U(1)_{2+1}$ lattice gauge theory} 
The lattice action is given by~\cite{Loan03}
\begin{eqnarray}
\label{eq:SplitGaugeAction}
S[U] &=& \frac{1}{g^{2}}\frac{a}{a_{0}} \sum_{\Box_{time-like}} 
[1 -Re(U_{\Box})]  \nonumber \\
&+& \frac{1}{g^{2}}\frac{a_{0}}{a} \sum_{\Box_{space-like}} 
[1 -Re(U_{\Box})] \nonumber \\
&\equiv&  S_{elec}[U] + S_{magn}[U] ~ .
\end{eqnarray}
The corresponding lattice Hamiltonian is given by~\cite{Irving83}
\begin{eqnarray}
\label{eq:GaugeHamilton}
H &=& \frac{g^{2}}{2a} \sum_{<ij>} \hat{l}_{ij}^{2} + \frac{1}{g^{2}a} \sum_{\Box_{space-like}} [1 - Re(U_{\Box})] \nonumber \\
&\equiv& H_{elec} + H_{magn} ~ ,
\end{eqnarray}
representing the electric term and a magnetic term, respectively. 
The electric term is built from the operator $\hat{l}_{ij}$ which represents the electric flux strings. 
Its eigenstates are 
\begin{equation}
\label{eq:ElectField}
\hat{l}_{ij} |\lambda_{ij}> = \lambda_{ij} |\lambda_{ij}> ~ , \lambda_{ij} = 0,\pm 1, \pm 2, \dots ~ .
\end{equation}
For each link $ij$, the states $|\lambda \rangle$ form a complete orthogonal basis,
\begin{equation}
\sum_{\lambda = 0,\pm1,\pm2,\dots} | \lambda \rangle \langle \lambda | = 1 ~ , ~
\langle \lambda' | \lambda \rangle = \delta_{\lambda',\lambda} ~ .
\end{equation}
The magnetic term is built from link operators $\hat{U}_{ij}$.
It has the eigenstates
\begin{equation}
\hat{U}_{ij} | U_{ij} \rangle = {U}_{ij} | U_{ij} \rangle ~ .
\end{equation}
For each link $ij$, this basis is a complete orthogonal basis,
\begin{equation}
\int dU ~ |U \rangle \langle U | = 1 ~ , ~
\langle U' | U \rangle = \delta(U'-U) ~ .
\end{equation}
The transition from the flux string basis to the link basis 
is determined by the commutator $[\hat{l}, \hat{U}] = - \hat{U}$ 
and yields~\cite{Paradis07} 
\begin{equation}
\label{eq:ScalarProd}
<\lambda|U> = (U)^{\lambda} ~ .
\end{equation}
We construct Euclidean transition matrix elements
\begin{eqnarray}
\label{eq:TransAmplGaugeInv}
M_{U_{f},U_{i}} 
&=&
\langle U_{f} | \hat{\Pi} \exp[-H T] | U_{i} \rangle \nonumber \\
&=& \left. \int [dU] \exp[- S[U] ] \right|^{U_{f},T}_{U_{i},0} ~ ,
\nonumber \\
\end{eqnarray}
where the operator $\hat{\Pi}$ (commuting with the Hamiltonian) denotes a projection operator states onto gauge invariant states. 
\begin{figure}[h,t]
\centering
\includegraphics[width=0.49\textwidth]{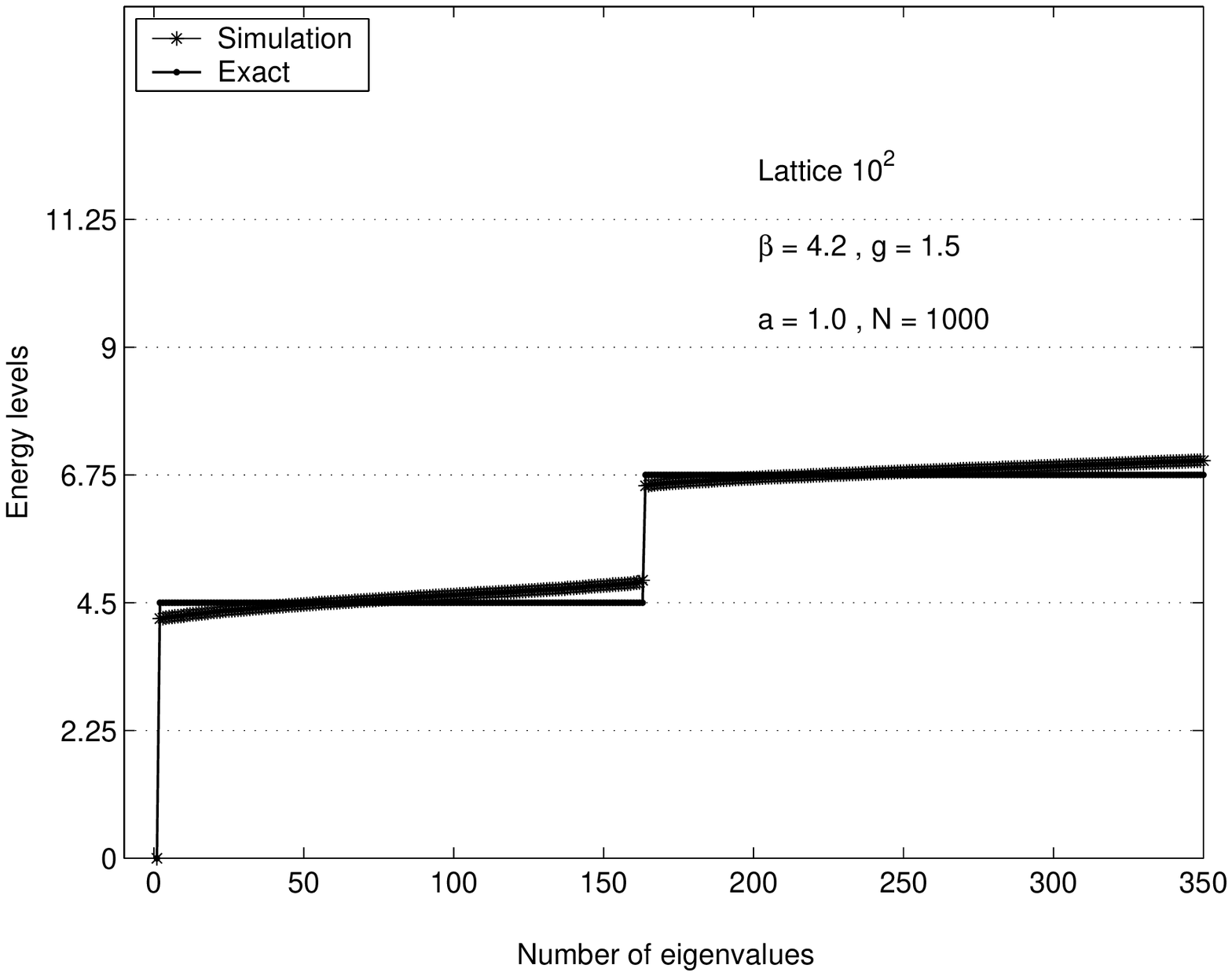}
\includegraphics[width=0.49\textwidth]{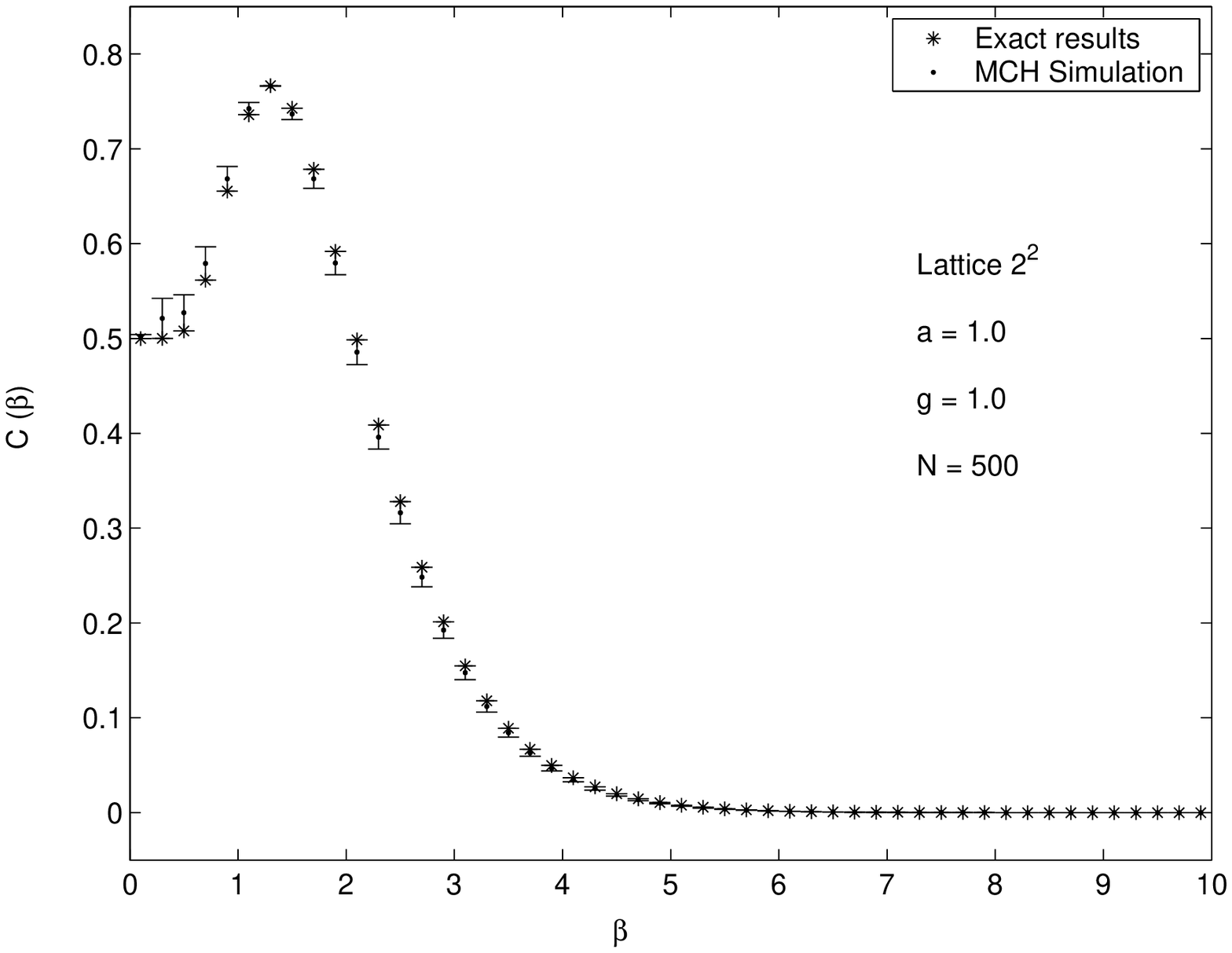} 
\caption{ 
Electric part of Hamiltonian. Comparison of exact results with those from Monte Carlo Hamiltonian.  
Left: Low-lying spectrum. $10^{2}$ spatial lattice, $a_{s}=a_{t}=1$, $g=1.5$, $T=4.2$, $N=1000$.
Right: Specific heat $C(\beta)$, $2^{2}$ spatial lattice, 
$a_{s}=a_{t}=1$, $g=1$, $N=500$. Error bars indicate statistical errors.
}
\label{fig:Spec4+9+100Plaq}
\end{figure}

\section{Stochastic basis}
For a system with a given Hamiltonian $H$ the physically motivated choice for the distribution is given by the transition amplitude in imaginary time, involving the same Hamiltonian,
\begin{equation}
\label{eq:TrialProbDistr}
P(U) = \langle U | \exp[-H T/\hbar ] | U_{init} \rangle ~ , 
\end{equation}
where $U_{init}$ is some suitably chosen fixed spatial lattice configuration (Bargmann state). This function is suitable as probability distribution because it is a positive function $P(U) \ge 0$. Physically relevant configurations (Bargmann states) can be drawn from this distribution by expression $P(U)$ as path integral and doing the sampling via Monte Carlo.
As alternative one may consider the distribution given by the transition amplitude from the electric Hamiltonian,
\begin{equation}
\label{eq:DefProbDistr}
P(U) = \langle U | \exp[-H_{elec} T/\hbar ] | U_{init} \rangle ~ , 
\end{equation}
which is an analytically computable function. 
Both of the above distributions involve a time parameter $T$, which determines the "width" of the distribution. Such time parameter needs to be tuned. As a general rule, we used to choose $T$ such that 
the energy spectrum falls into a scaling window of eigenvalues (see below).

\section{Transition amplitudes from electric Hamiltonian}
\label{sec:AnalExpr}
Applying the Peter-Weyl theorem~\cite{Tung} to the group $U(1)$ allows to expand link states in terms of 
irreducible representation matrices. In the case of the group $U(1)$, the Peter-Weyl theorem is equivalent to Fourier expansion~\cite{Tung}. The Peter-Weyl theorem holds more generally for groups $SU(N)$~\cite{Vilenkin}. 
Applying the Peter-Weyl theorem for the group $U(1)$ to the transition amplitude between two single link states reads
\begin{eqnarray}
\label{eq:PeterWeylU1}
&&<U_{fi}|\exp[- H_{elec} T/\hbar ]| U_{in}>  \nonumber \\ 
&=& \sum_{n=0,\pm1,\pm2,\dots} 
\exp[- \frac{g^{2} \hbar T}{2a} n^{2} ]
\cos[n(\alpha_{in} - \alpha_{fi})] ~ . 
\nonumber \\
\end{eqnarray}
In mathematical terms $n$ (running over $0,\pm1,\pm2,\dots$) denotes the index 
of the irreducible representation. $(U)^{n}$ denotes the irreducible representation 
of group element $U$ with representation index (quantum number) $n$. In physical 
terms, $n$ represents the number of electric flux lines. 
The Hamiltonian $H_{elec}$ is a Casimir, which is diagonal in the representation index 
$n$. The link variables have been parametrized via $U = \exp[i \alpha]$.

The Peter-Weyl theorem is also useful for the construction of gauge invariant states. 
For example, let us take a spatial lattice consisting of four links ordered to form a plaquette and consider the transition amplitude between initial and final Bargmann states. 
In order to make the amplitude gauge invariant, we carry out the group integral over the gauge orbit at each node. The group integral generates Gauss' law enforcing conservation of the number of flux lines at each vertex.
By defining the plaquette angle 
$\theta_{plaq} = \alpha_{12} + \alpha_{23} + \alpha_{34} + \alpha_{41}$
and $\Delta \theta_{plaq} = \theta^{fi}_{plaq}  - \theta^{in}_{plaq}$, 
we obtain the final expression of the gauge invariant amplitude, 
\begin{eqnarray}
\label{eq:1PlaqAmplGaugeInv2}
&&\langle U^{fi} | \hat{\Pi} \exp[-H_{elec}T/\hbar] | U^{in} \rangle \nonumber \\
&=&
\sum_{n=0,\pm1,\pm2,\dots} \exp \left[ - \frac{g^{2}\hbar T}{2a} 4 n^{2} \right] ~ 
\cos \Bigl[ n \Delta \theta_{plaq}  \Bigr] ~ .
\nonumber \\
\end{eqnarray}
Here $n$ denotes the number of closed plaquette loops. The result is built from plaquettes which are closed loops of consecutive link variables forming the smallest non-local gauge invariant objects on the lattice. The eigenvalue of the electric field $\vec{E}^{2}$ corresponds to the contribution from $n$ plaquette loops (on top of each other). The result only depends on the number of plaquette loops and the difference between initial and final plaquette angles.

\section{Test of Monte Carlo Hamiltonian}
A. {\it Electric Hamiltonian}. 
In order to test the Monte Carlo Hamiltonian, we first consider the electric part of the Hamiltonian. 
This is a good test bed, because the spectrum of the electric Hamiltonian 
can be computed analytically.
Fig.[\ref{fig:ScalingCompareElec+FullHam}](left) shows the low-lying part of the spectrum as function of the time parameter $\beta=T$ occuring in the matrix elements. The physical spectrum should be independent of the time parameter. In the numerical results this is reflected by the existence of scaling windows (region of flat line). The size of such scaling window depends on the particular energy-level and decreases with increasing energy.
We found that the size of the scaling window $S_{n}$ can be described approximately 
by an exponential law $S_{n} \propto \exp[-\sigma E_{n}]$. Such behavior of decreasing 
scaling windows can be understood from the property that $\exp[-HT/\hbar]$ projects 
onto the ground state for large $T$ (Feynman-Kac theorem). Higher levels become 
exponentially suppressed by the dominant ground state and can survive only for 
short times $T$. 
We looked also for scaling windows in the corresponding wave functions.
In particular, we have studied 
$\langle e_{\mu} | \Phi_{n} \rangle$, i.e. the expansion coefficient of wave 
function $\Phi_{n}$ in terms of the stochastic basis function $e_{\mu}$. Such 
expansion coefficients for the first energy levels expanded in terms of the 
first basis function also display scaling windows (not shown).  
Like the size of the energy scaling window decreases with increasing energy $E_{n}$,
also the size of the wave function scaling window decreases with the 
level index $n$ of energy.

For the case of low-lying spectrum using a lattice of size $10^{2}$ a comparison of results from the Monte Carlo Hamiltonian with 
the exact spectrum is shown in Fig.[\ref{fig:Spec4+9+100Plaq}](left). 
The high degeneracy is due to 1- and 2-plaquette states located anywhere on the $10^{2}$ 
lattice (such degeneracy will be lifted when taking the magnetic term into account). The 
figure shows, firstly, that the Monte Carlo Hamiltonian captures almost all of the degenerate 
states and secondly, reproduces the exact energies with small error. 
The quality of the energy spectrum of the Monte Carlo Hamiltonian can be seen also 
from a look at thermodynamical functions. In the case of the electric Hamiltonian, the energy 
spectrum $E_{0}, E_{1},\dots$ can be computed analytically. Thermodynamical functions 
can be expressed via those energies. 
We have computed average energy, free energy,  
entropy and specific heat,  
and compared the results from the Monte Carlo Hamiltonian with the exact one. The result for the specific heat is displayed in 
Fig.[\ref{fig:Spec4+9+100Plaq}](right).
In general, one observes very good agreement in the regime of large $\beta$, 
i.e., the low temperature regime. For small values of $\beta$ some disagreement 
becomes visible, reflecting the fact that the precision of higher enegy levels 
of the Monte Carlo Hamiltonian is limited (their scaling windows go to zero).
\\

B. {\it Including the magnetic term: full Hamiltonian}.
Finally, we consider the gauge invariant transition amplitude under the full Hamiltonian (Eq.\ref{eq:TransAmplUpsilon}). Although it can be expressed in terms of a path integral with the lattice action (Wilson action), this is numerically not suitable, because Monte Carlo with importance sampling only allows to compute ratios of transition amplitudes. Hence, we factorize the above amplitude into two terms, one being analytically computable and the other one being given by the ratio of transition amplitudes computable via Monte Carlo, 
\begin{eqnarray}
\label{eq:MatrixElemGauge}
M_{\mu,\nu}(T) 
&=& \langle U_{\mu} |\hat{\Pi} ~ \exp[-H_{elec} T/\hbar ] | U_{\nu} \rangle 
\nonumber \\
&\times& \frac{ \langle U_{\mu} |\hat{\Pi} ~ \exp[-H T/\hbar ] | U_{\nu} \rangle }
{ \langle U_{\mu} |\hat{\Pi} ~ \exp[-H_{elec} T/\hbar ] | U_{\nu} \rangle }
~ .
\end{eqnarray}
Taking into account the magnetic term allows to obtain the full Monte Carlo Hamiltonian.
First results on scaling of its low-lying energy spectrum are shown in 
Fig.~[\ref{fig:ScalingCompareElec+FullHam}](right). These results correspond to a 
small lattice ($2^{2}$) and also a small number of basis functions. The results show scaling windows for the lowest five energy levels. 
Compared to the scaling behavior observed in the electric Hamiltonian 
(Fig.~[\ref{fig:ScalingCompareElec+FullHam}]~left) fewer levels show scaling, 
and the scaling windows are smaller. This can be understood from the fact that 
the ratio of matrix elements in Eq.~(\ref{eq:MatrixElemGauge}) has been determined 
via Monte Carlo from path integrals, which carries statistical errors in the order 
of a few percent. 
From this observation we conclude that the numerical resolution of 
energy levels and the size of scaling windows of the full Hamiltonian is essentially determined 
by the statistical error occuring in the numerical calculation of the ratio of 
matrix elements. Results with better statistics, a larger stochastic basis and larger lattice volumes are 
under way.

\vspace*{6pt}

\noindent {\bf Acknowledgement.} H. Kr\"oger has been supported by NSERC Canada. This paper is dedicated to the memory of Prof. X.Q. Luo.

\end{document}